\documentclass[12pt]{iopart}
\usepackage{graphics}
\usepackage{epsfig}
\usepackage{citesort}

\usepackage{latexsym}
\usepackage{epsfig}
\usepackage{amssymb}
\usepackage{latexsym}
\usepackage{epsfig}
\usepackage{amssymb}

\begin{document}

\title{Probing Dark Energy at Galactic and Cluster Scales}
\author{David F. Mota}
\mailto{D.Mota@thphys.uni-heidelberg.de}
\address{Institute for Theoretical Physics, University of Heidelberg, 69120 Heidelberg,Germany}
\address{Lab. de Physique Th´eorique et Astroparticules, CNRS Universit´e Montpellier II, France}

\begin{abstract}
We investigate dark matter halo properties as a function of a time--varying dark energy equation
of state. The dynamics of the collapse of the halo 
is governed by the form of the quintessence potential, the time evolution of its 
equation of state, the initial conditions of the field and its homogeneity
nature in the highly non--linear regime. These have a direct
impact on the turnaround, virialisation and collapse times,
altering in consequence the non--linear density contrast and virial radius. 
We compute halo concentrations using the Eke, Navarro \& 
Steinmetz algorithm, examining two extreme scenarios: first, we assume that
the quintessence field does not exhibit fluctuations on cluster scales
and below - homogeneous fluid; second, we assume that the field inside the overdensity 
collapses along with the dark matter - inhomogeneous fluid. 
The Eke, Navarro \& Steinmetz prescription reveals, in general, higher halo concentrations 
in inhomogeneous dark energy models than in their homogeneous equivalents. 
Halo concentrations appear to be controlled by both changes in formation epochs of the 
halo cores as well as by differing virialisation overdensities.
We derive physical halo properties in all models and discuss their observational 
implications.  We examine two possible methods for comparing observations 
with theoretical predictions. The first method works on galaxy cluster scales 
and consists of fitting the observed X--ray cluster gas density distributions to 
those predicted for an NFW profile. The second method works on galaxy scales and involves the observational measurement
 of the so--called central density parameter.
\end{abstract}


\maketitle

\section{Introduction}

Observations of the Cosmic Microwave Background temperature anisotropy reveal 
that a mysterious constituent with negative pressure, so--called dark energy, 
accounts for 70 percent of today's mass--energy budget and is causing the expansion of the 
universe to accelerate \cite{spergel}. These observations are in remarkable 
concord with the observations of distant supernovae \cite{perl}. 

The present--day challenge in cosmology is to discover the physical nature of 
dark energy. Candidates for this unknown entity include the cosmological constant, a dynamic cosmic field, commonly designated 
quintessence, which varies across space and changes with time, and a possible modification of General Relativity at large
scales \cite{carroll,ratra,mod}. 

Although the physical nature of dark energy is undisclosed, one can explore its 
effects on cosmic structure formation, in particular one can study its implications 
for the number density of dark matter halos and their density profiles. In this 
respect, significant progress has been made by several authors who performed numerical 
N--body simulations in dark energy models \cite{klypin,dolag,kuhlen,o1,o2}. These investigations 
are imperative for cosmological studies that rely on these ingredients to measure dark 
energy. Examples of this studies include semi-analytical studies of strong lensing 
statistics \cite{bart03,lopes,o3} and weak lensing number counts \cite{bart03,manera}.

In this paper, we investigate how halo properties change in cosmological models 
with dynamical dark energy. This work extends upon previous studies in that we 
examine halo properties as a function of a time--varying dark energy equation 
of state, covering four types of potentials, and its homogeneity nature in the 
highly non--linear regime. We utilise the predictions of the spherical collapse model, 
such as the virial overdensity, obtained by \cite{mota,maor}, and calculate halo 
concentrations using the semi-analytical algorithm of  Eke, Navarro \& Steinmetz (ENS) \cite{eke}. We then derive 
physical halo properties in all models and discuss their observational implications.

The behaviour of linear perturbations in a scalar field and its effect on 
structure formation has been investigated by a number of authors. However, the behaviour 
of quintessence during the non--linear gravitational collapse is not well understood and 
is currently under investigation (see e.g. \cite{mota,maor,perrota,wetterich02,amendola} for recent work).
Usually, it is assumed that the quintessence field does not exhibit density fluctuations 
on cluster scales and below. The reason for this assumption is that, according to
linear perturbation theory, the mass of the field is very small (the associated
wavelength of the particle is of the order of the Hubble radius) and, hence, it
does not feel matter overdensities of the size of tenth of a Mpc or smaller \cite{wang}.

The assumption of neglecting the effects of matter perturbations on the
evolution of dark energy at small scales is indeed a good
approximation when perturbations in the metric are very small. However, care must be taken 
when extrapolating the small--scale linear--regime results to the highly non--linear regime. 
Then, locally the flat FRW metric is no longer a good approximation to describe the geometry
 of overdense regions. 
Highly non--linear matter 
perturbations could, in principle, modify the evolution of perturbations in dark energy 
considerably, and these could, in turn, backreact and affect the evolution of matter 
overdensities. 
Moreover, it is natural to think that once
a dark matter overdensity decouples from the background expansion and collapses, the field
inside the cluster feels the gravitational potential inside the overdensity and  
its evolution will be different from the background evolution. 
This is a general feature of many 
cosmological scalar fields whose properties depend on
the local density of the region they ``live in''
\cite{mota1}.

\cite{bean} suggested that the quintessence field could have an important impact in the 
highly non---linear regime. \cite{wetterich02,arbey} noted
that the quintessence field could indeed be important on galactic
scales. It was put forward by \cite{guz} that it could in fact
be responsible for the observed flat rotation curves in
galaxies. Other authors \cite{pad1} discussed more
exotic models, based on tachyon fields, and argued that the equation
of state is scale--dependent.

If it turns out that the effects of dark matter density perturbations and metric  
influence perturbations of quintessence on small scales, this could significantly change 
our understanding of structure formation on galactic and cluster scales.
\cite{mota,maor,rogerio} have shown that properties of halos, such as the 
density contrast and the virial radius, depend critically on the form of the potential, the
initial conditions of the field, the time evolution of its equation of state and on the 
behaviour of quintessence in highly non--linear regions.
In reality, the dependence on the inhomogeneity of dark energy is only important for some 
dark energy candidates. If the dark energy equation of state is 
constant, the differences between the homogeneous and inhomogeneous cases are 
small, as long as the equation of state $w$ does not differ largely from $w=-1$ 
\cite{mota,maor}. Thus, for constant equation of state, the fitting
formulae for the cold dark matter (CDM) density contrast presented in
the literature \cite{wang,weinberg}, do not  
change drastically, even if inhomogeneities in the dark energy component are taken into 
account.  

The paper is organised as follows. In section 2 we describe briefly the 
spherical collapse model and its dependence on the homogeneity nature of dark energy. 
In section 3 we calculate halo concentrations in both homogeneous and inhomogeneous 
dark energy models. We diagnose observational methods to measure physical halo properties in 
section 4. Section 5 discusses the results and draws the conclusions.

\section{The spherical collapse model and the homogeneity nature of dark energy}

We consider a flat, homogeneous and isotropic background universe with 
scale factor $a(t)$. Since we are interested in the matter dominated epoch, 
when structure formation starts, we fill the universe with cold
dark matter of density $\rho_{\rm m}$ $\propto a^{-3}$ and a 
dark energy fluid with energy density $\rho_{\phi}$. 
The equations that describe the background universe
are (we set $\hbar =c\equiv 1$ throughout the paper):
\begin{eqnarray}
3H^{2}&=&8\pi G\left( \rho_{\rm m}+\rho_{\phi }\right)  
\label{fried}\\
\dot\rho_{\phi}&=&-3 H (1+w_{\phi})\rho_{\phi}
\label{psidot}
\end{eqnarray}
where $H\equiv \dot{a}/a$ is the Hubble rate. 
When $w_{\phi}=-1$ dark energy is the
vacuum energy density.
If dark energy is a scalar field $\phi$ (quintessence), 
$\rho_{\phi}=\frac{1}{2}\dot\phi^2+V(\phi)$ and
$P_{\phi}=\frac{1}{2}\dot\phi^2-V(\phi)$, where $V(\phi)$ is the
scalar field potential. In this case, it is useful to rewrite
equation (\ref{psidot}) as
\begin{equation}
\ddot\phi+3 H \dot\phi+V^{'}=0
\label{phidot}
\end{equation}
where the prime represents a derivative with respect to $\phi$.

\begin{figure}
\resizebox{100mm}{!}{
\rotatebox{-90}{
\includegraphics{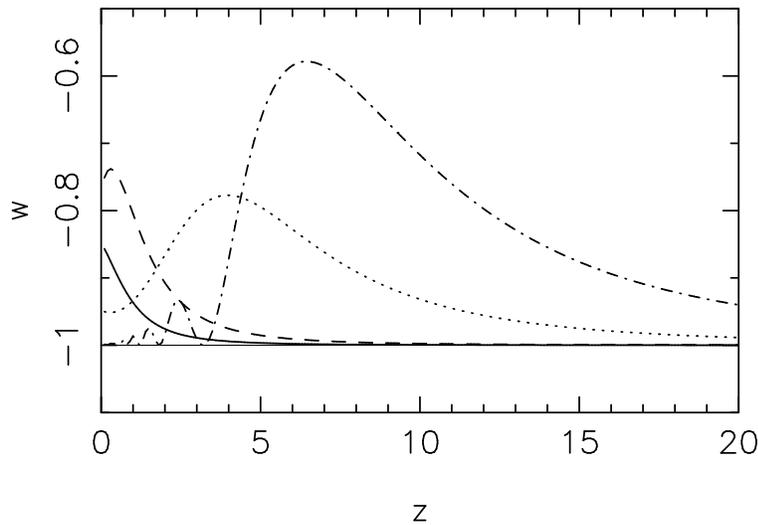}}}
\caption{Dark energy equation of state in the background universe as a function of 
redshift, $z$. Thick solid line - \cite{copel}, dashed line - \cite{steinhardt}, 
dashed--dotted line - \cite{skordis}, dotted line - \cite{brax}, thin solid line - 
cosmological constant.}
\label{wz}
\end{figure}

In this paper, we consider four examples of quintessential potentials:
\begin{itemize}
\item The double exponential potential \cite{copel}:
\begin{equation}
V(\phi) = M\left(\exp(\beta\phi)+\exp(\gamma\phi)\right).
\end{equation}
\item The exponential potential with inverse power \cite{steinhardt}:
\begin{equation}
V(\phi) = M\left(\exp(\gamma/\phi)-1\right)
\end{equation}
\item The Albrecht--Skordis potential \cite{skordis}:
\begin{equation}
V(\phi) = M\left(A + \left(\phi - B\right)^2\right)\exp(-\gamma\phi)
\end{equation}
\item The supergravity--motivated potential \cite{brax}:
\begin{equation}
V(\phi) = M\exp(\phi^2)/\phi^\gamma
\end{equation}
\end{itemize}
We choose the parameters in the potentials and the initial conditions in 
the background such that we obtain the following present--day cosmological 
parameters: $\Omega_m=0.3$, $\Omega_{\phi}=0.7$,$H_{0} = 100$\,h\,km\,s$^{-1}$\,Mpc$^{-1}$ 
with $h=0.7$, and $-1\leq w_{\phi}\leq-0.8$.
The background time evolution of the equation of state for each model is shown in figure \ref{wz}.

The evolution of a spherical overdense
patch of scale radius $R(t)$ is given by the Raychaudhuri equation: 
\begin{equation}
3\ddot{R}=-4\pi G R\left( \rho _{\rm m_{c}}+\rho _{\phi_{c}}(1+3w_{\phi_{c}})\right)  
\label{rcluster}
\end{equation}
Note that it would be wrong to use the Friedmann equation
for a closed universe with a constant curvature $k$, since the former can 
vary in time \cite{wang,weinberg}.
In the halo, the evolution of $\rho_{\phi _{c}}$ and $\rho _{\rm m_{c}}$ 
is given by 
\begin{equation}
\dot\rho_{\phi_{c}}=-3\frac{\dot R}{R}(1+w_{\phi_{c}})\rho_{\phi_{c}} + \Gamma
\label{psidotcluster}
\end{equation}
and $\rho _{\rm m_{c}}\propto R^{-3}$ due to mass conservation. 
Once again, in the case of a scalar field, equation (\ref{psidotcluster}) 
can be written as \cite{mota,maor}  
\begin{equation}
\ddot{\phi_{c}}+3\frac{\dot{R}}{R}\dot{\phi_{c}}+V_c{'}(\phi_{c})=
\frac{\Gamma}{\dot\phi_{c}} 
\label{phidotcluster}
\end{equation}
where $\phi_c$ is the field inside the overdensity (the lowerscript distinguishes it 
from the background value) and $V_{c}=V(\phi_{c})$ is its potential.
The quantity $\Gamma$ describes the energy loss of dark energy inside the dark matter 
overdensity, as this component does not necessarily follow the collapse of 
dark matter, and energy can formally flow out of the system. As such, $\Gamma$ encodes 
in how far dark matter non--linearities act on the dark energy component. 

We will make two assumptions for $\Gamma$. In the first case, 
the quintessence field is assumed to be smooth throughout space. 
This corresponds to
\begin{equation}
\Gamma = -3\left(\frac{\dot a}{a}-\frac{\dot R}{R}\right)\dot\phi^2_{c}, 
\label{gamma}
\end{equation}
$\phi_{c}(t_i) = \phi(t_i)$ and $\dot\phi_{c}(t_i)=\dot\phi(t_i)$, which 
implies $\phi_{c}(t) = \phi(t)$ at all times, and is the standard
assumption made in the literature. 
In the second case, we assume that the field follows the dark matter collapse from the very 
beginning. That is, we assume $\Gamma = 0$.

Clearly, the values of $\Gamma$ chosen are not realistic, but they mark off two extreme 
scenarios.
At very late times, during the collapse of the dark matter, especially when the density 
contrast in dark matter is very large ($\delta_{\rm m}\gg1$), the field should no longer 
feel the background metric, i.e. it decouples from the background expansion. In this regime,
the evolution of dark energy could be different and influence the details of the collapse. 
Only when $\Gamma$ is quantified and the boundary conditions between the outer and 
inner metrics are properly understood, will the spherical collapse model be able make 
to solid predictions.
An estimate of $\Gamma$ can only be obtained from a general relativistic treatment. 
This could be achieved in two ways. One way would be to develop a 
swiss cheese model. Another way would be to perform N--body simulations that take into 
account the effects of dark energy perturbations in smaller scales. This is however out 
of the scope of this article.

Throughout the paper, we refer to a homogeneous dark energy model as a model where dark 
energy does not exhibit fluctuations on cluster scales and below
($\Gamma$ given by equation \ref{gamma}). This designation may be seen
as somewhat abusive in the sense that these models are inhomogeneous
on scales larger than cluster scales. In truth, only the cosmological
constant model is homogeneous across all space. We refer to a
inhomogeneous dark energy model as a model where dark energy does
exhibit fluctuations on cluster scales and below ($\Gamma=0$). The
terms homogeneous/inhomogeneous are therefore associated with cluster
scales and below, here taken as the scales of interest. 

We evolve the spherical overdensity from high redshift until its
virialisation occurs (see \cite{mota,maor} for details). Figure \ref{conc} plots the 
CDM density contrast, $\Delta_{\rm vir}\equiv \frac{\rho_{\rm m_{c}}}{\rho_{\rm crit}}$, as 
a function of redshift. 
It is clear from the figure that an homogeneous dark energy fluid is very similar to 
a cosmological constant. Major differences occur in the inhomogeneous case. These are 
discussed in the following section.

\section{Halo properties}

The dark matter halo is modelled with the NFW profile \cite{nfw}:
\begin{equation}
\rho(r)=\frac{\rho_{s}}{(r/r_{s})(1+r/r_{s})^{2}}
\end{equation} 
where $r_{s}$ is the characteristic length and $\rho_{s}$ is the characteristic
density scale, $\rho_{s}=\delta_{c}\rho_{\rm crit}$, where
$\rho_{\rm crit}=3H^{2}/8\pi G$ is the critical density for closure and  
\begin{equation}
\delta_{c}=\frac{\Delta_{\rm vir}}{3}\frac{c_{\rm vir}^{3}}{ln(1+c_{\rm vir})-c_{\rm vir}/(1+c_{\rm vir})}
\label{deltac}
\end{equation}
is the characteristic density contrast. The halo concentration is
defined as $c_{\rm vir}\equiv r_{\rm vir}/r_{s}$ where $r_{\rm vir}$ is the radius of a
sphere containing a mean density $\Delta_{\rm vir}$ times the critical
density:
\begin{equation}
M_{\rm vir}=\frac{4\pi}{3}  r_{\rm vir}^{3}\Delta_{\rm vir}\rho_{\rm crit}.
\label{mvir}
\end{equation} 
At small radii ($r \ll r_{s}$) $\rho(r) \sim r^{-1}$, and at large
radii ($r \gg r_{s}$) $\rho(r) \sim r^{-3}$. The NFW profile appears
to be a good fit to numerically simulated halos over a wide range of
masses in various cosmological scenarios. Although the exact inner
slope is under debate, there is general consensus that the density 
profile steepens at large radii. 

\subsection{Halo concentrations}

\begin{figure*}
\begin{minipage}{158mm}
\begin{center}
\resizebox{67mm}{!}{
\rotatebox{-90}{
\includegraphics{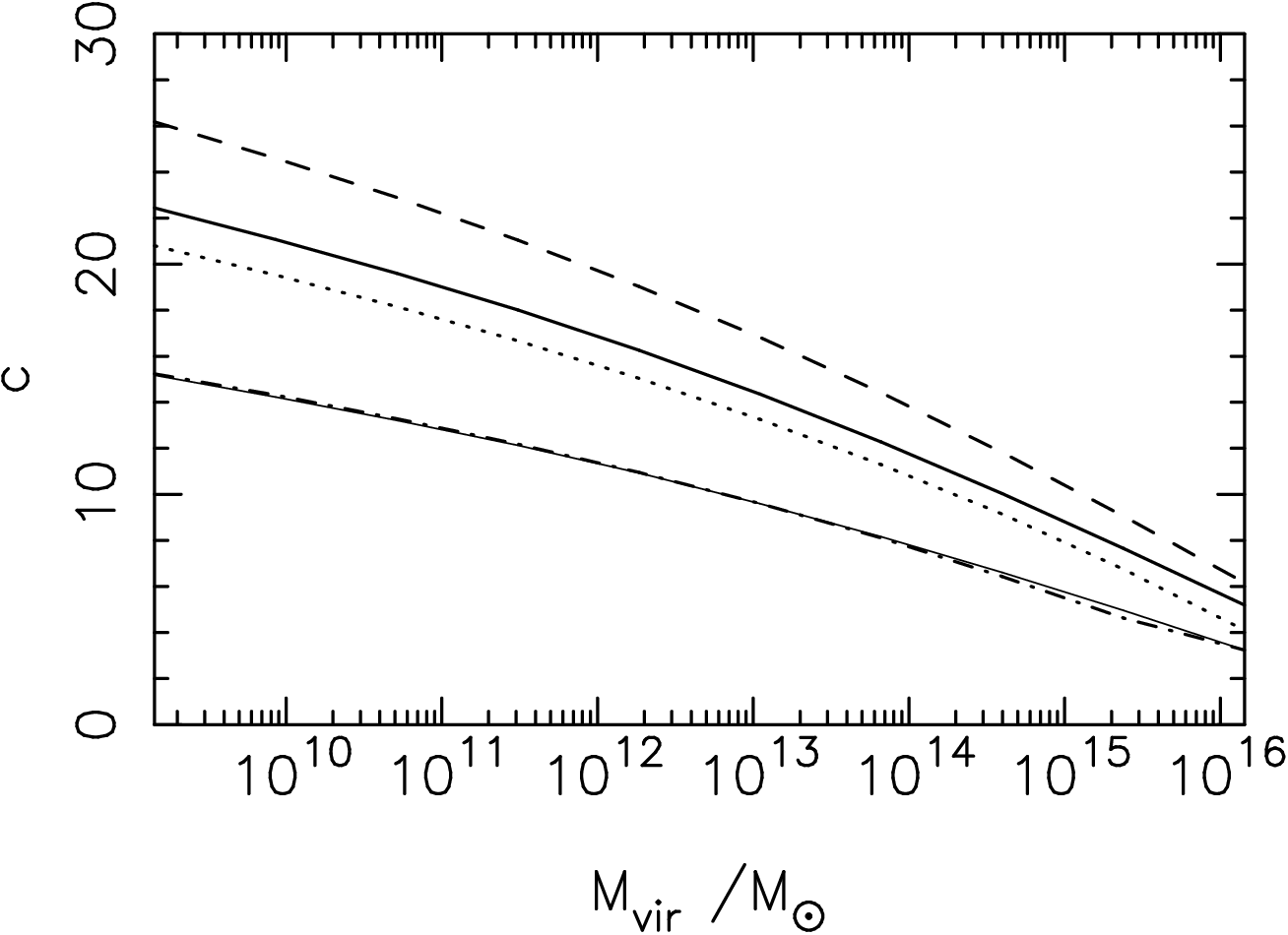}}
}
\resizebox{67mm}{!}{
\rotatebox{-90}{
\includegraphics{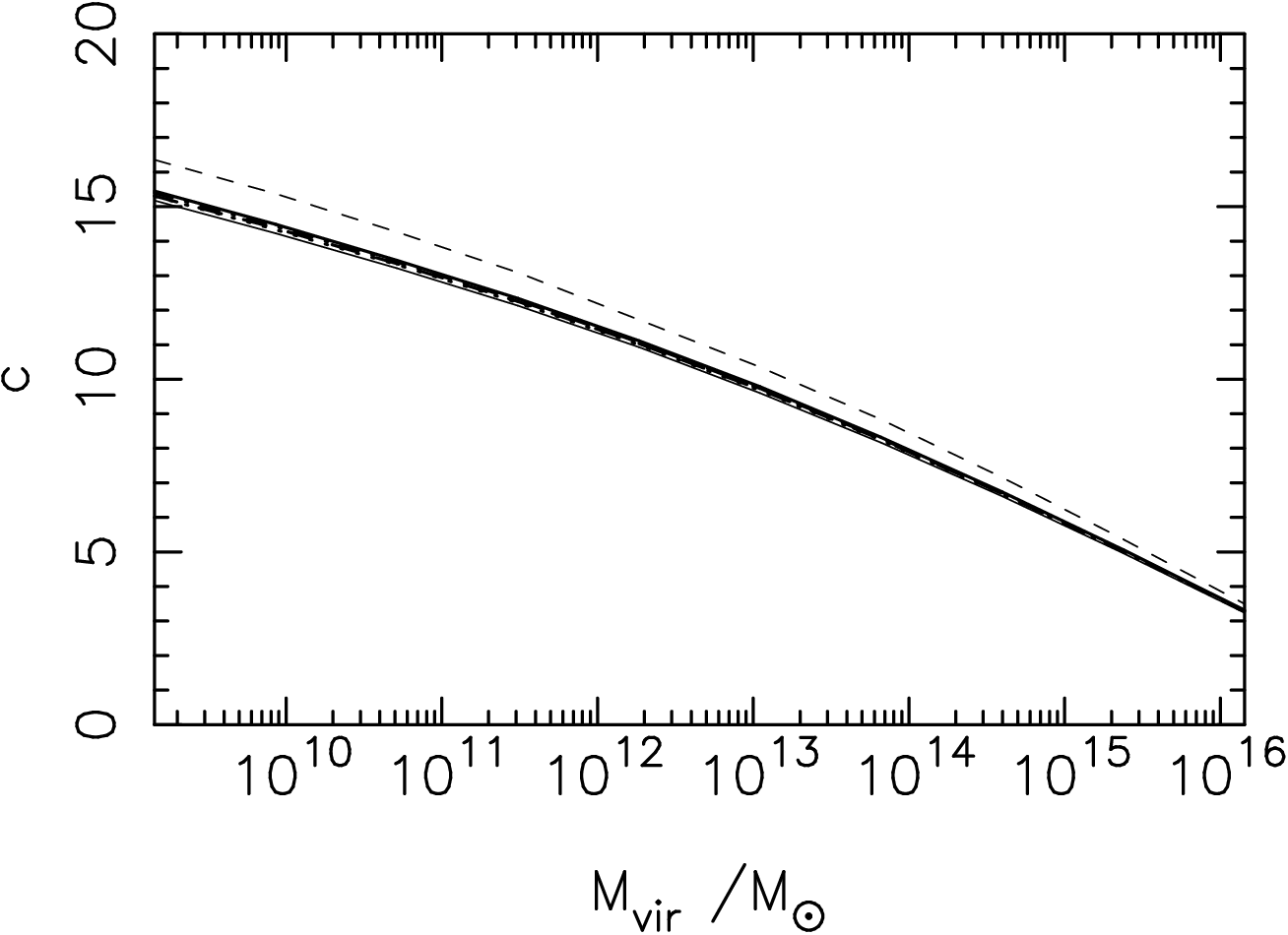}}
}

\resizebox{67mm}{!}{
\rotatebox{-90}{
\includegraphics{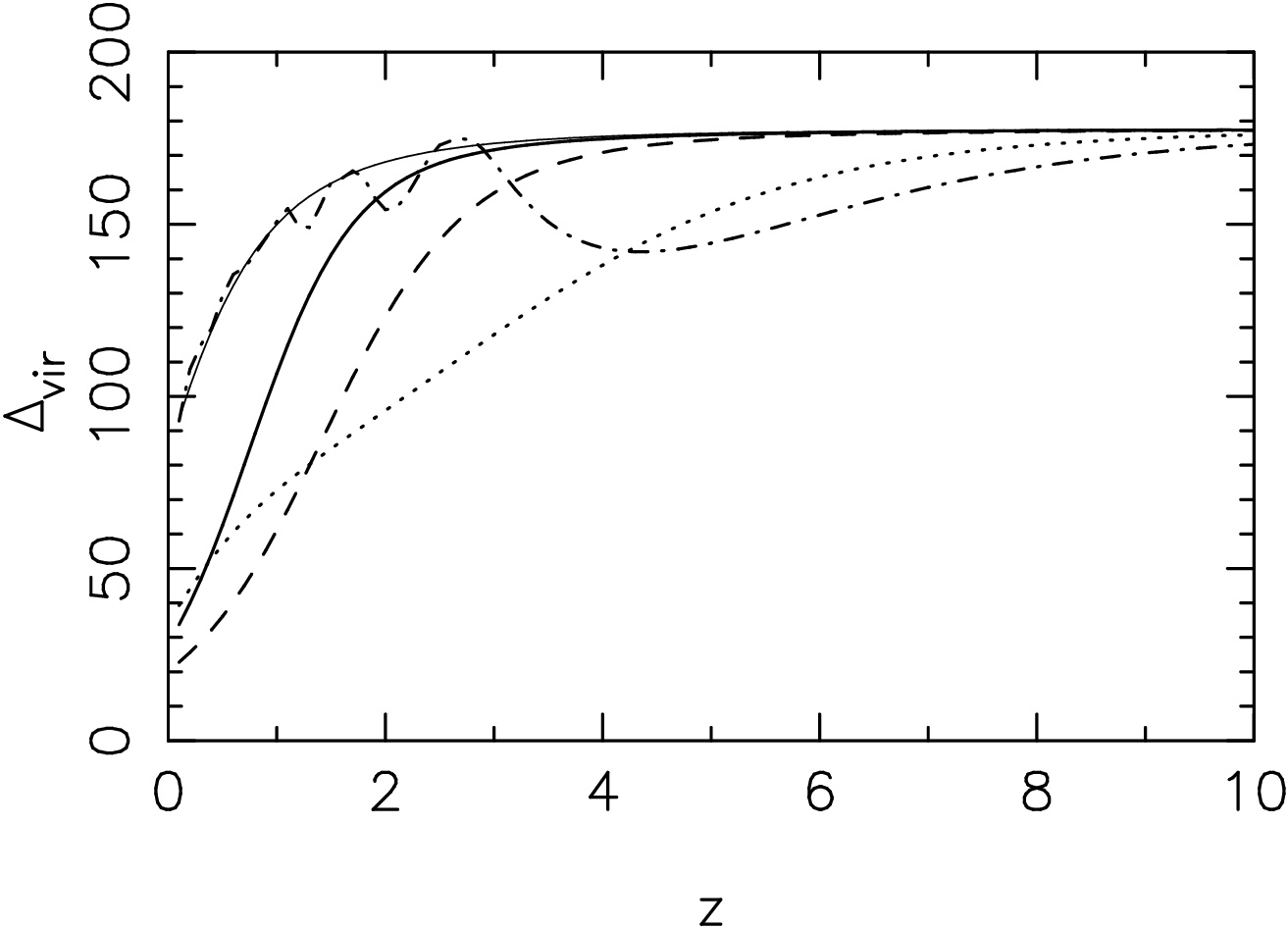}}}
\resizebox{67mm}{!}{
\rotatebox{-90}{
\includegraphics{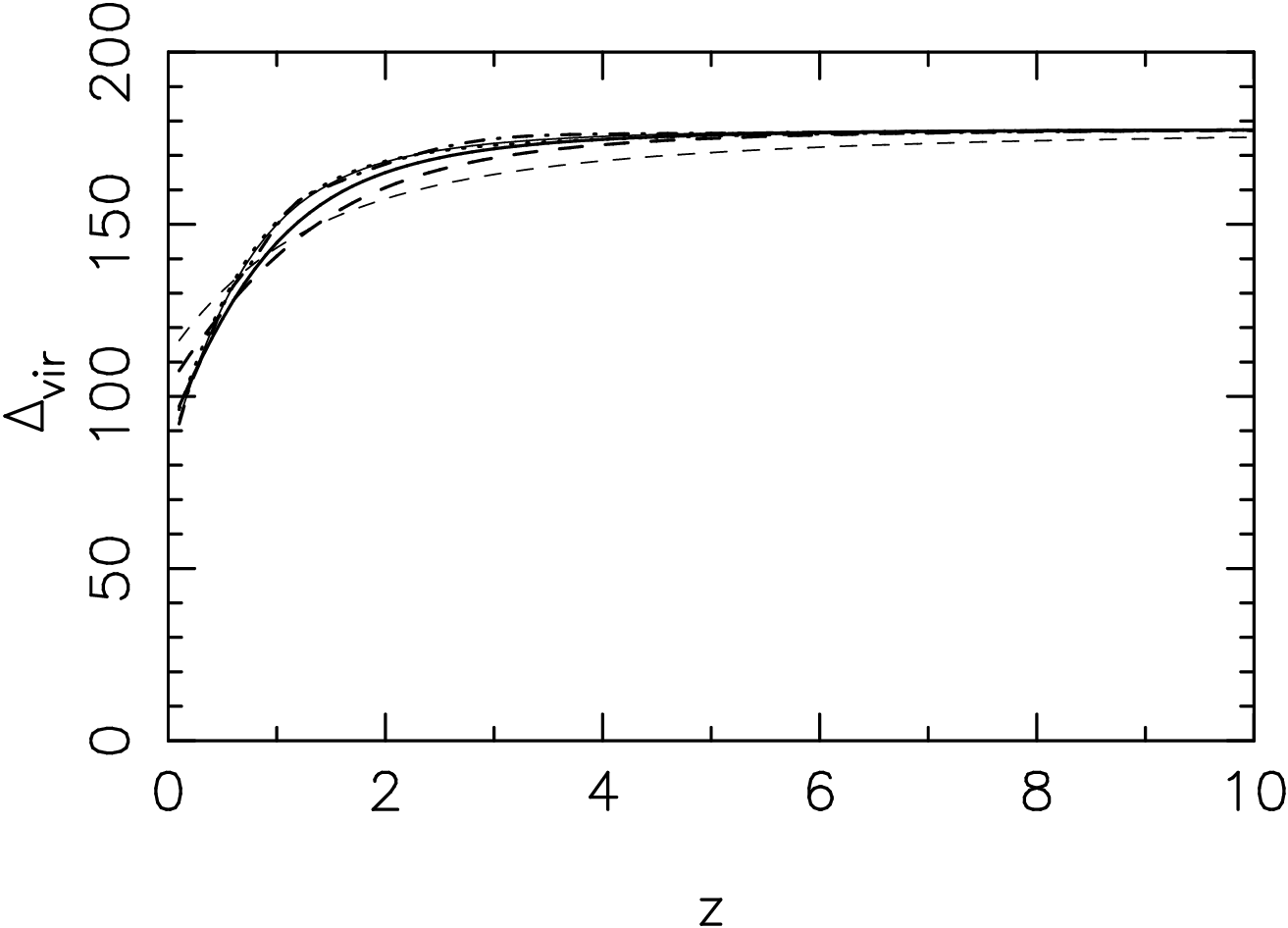}}}

\resizebox{67mm}{!}{
\rotatebox{-90}{
\includegraphics{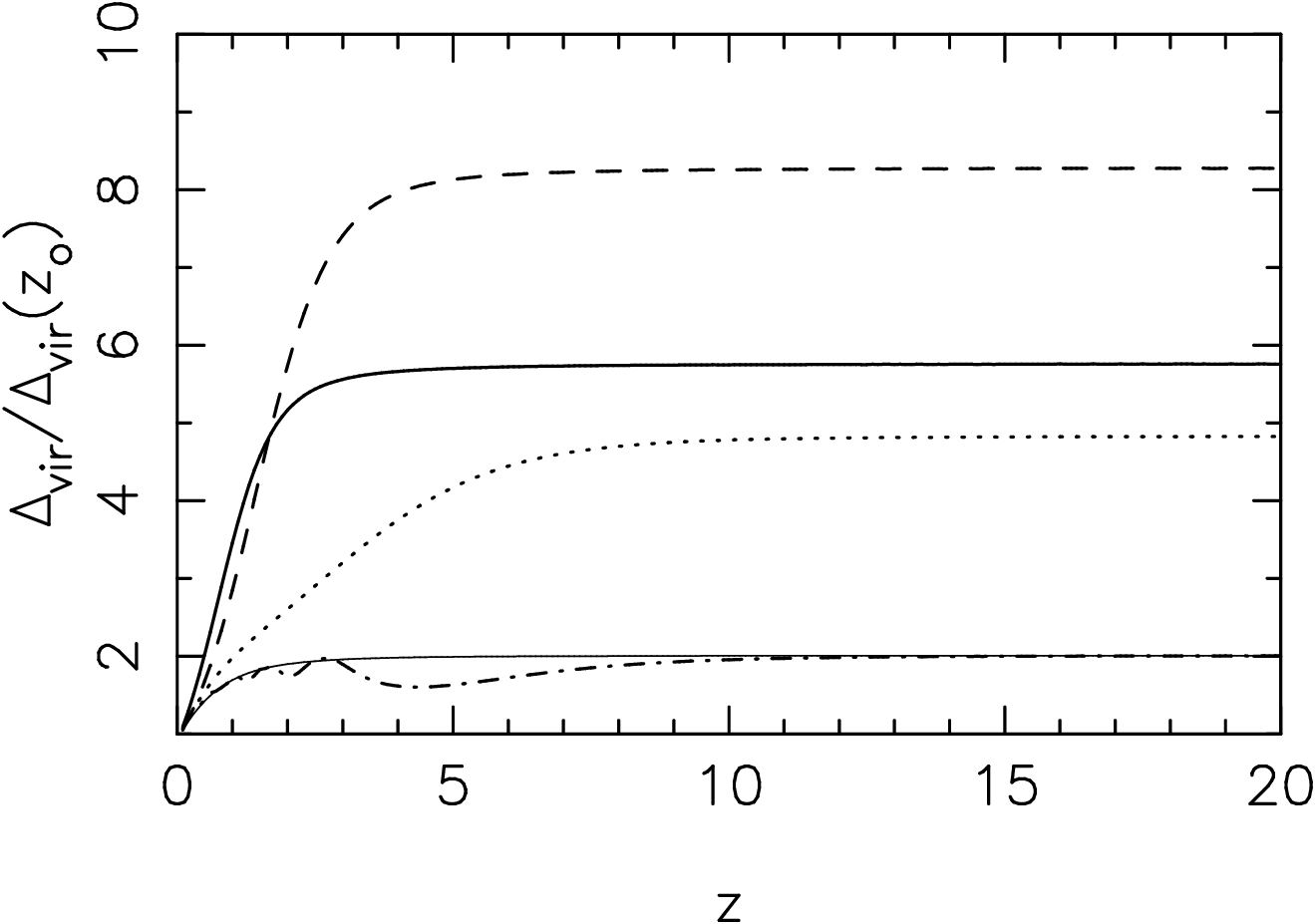}}}
\resizebox{67mm}{!}{
\rotatebox{-90}{
\includegraphics{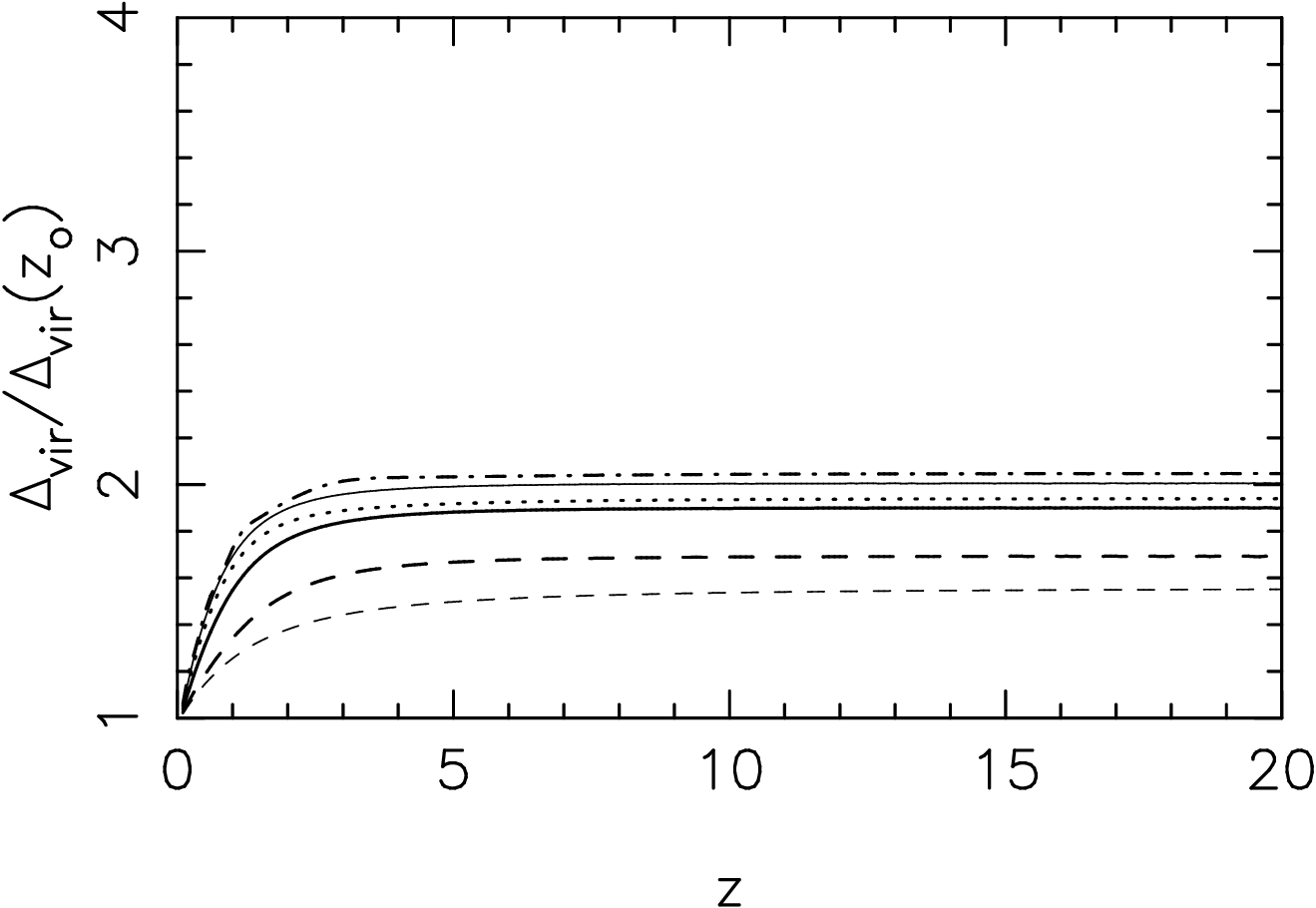}}}

\resizebox{67mm}{!}{
\rotatebox{-90}{
\includegraphics{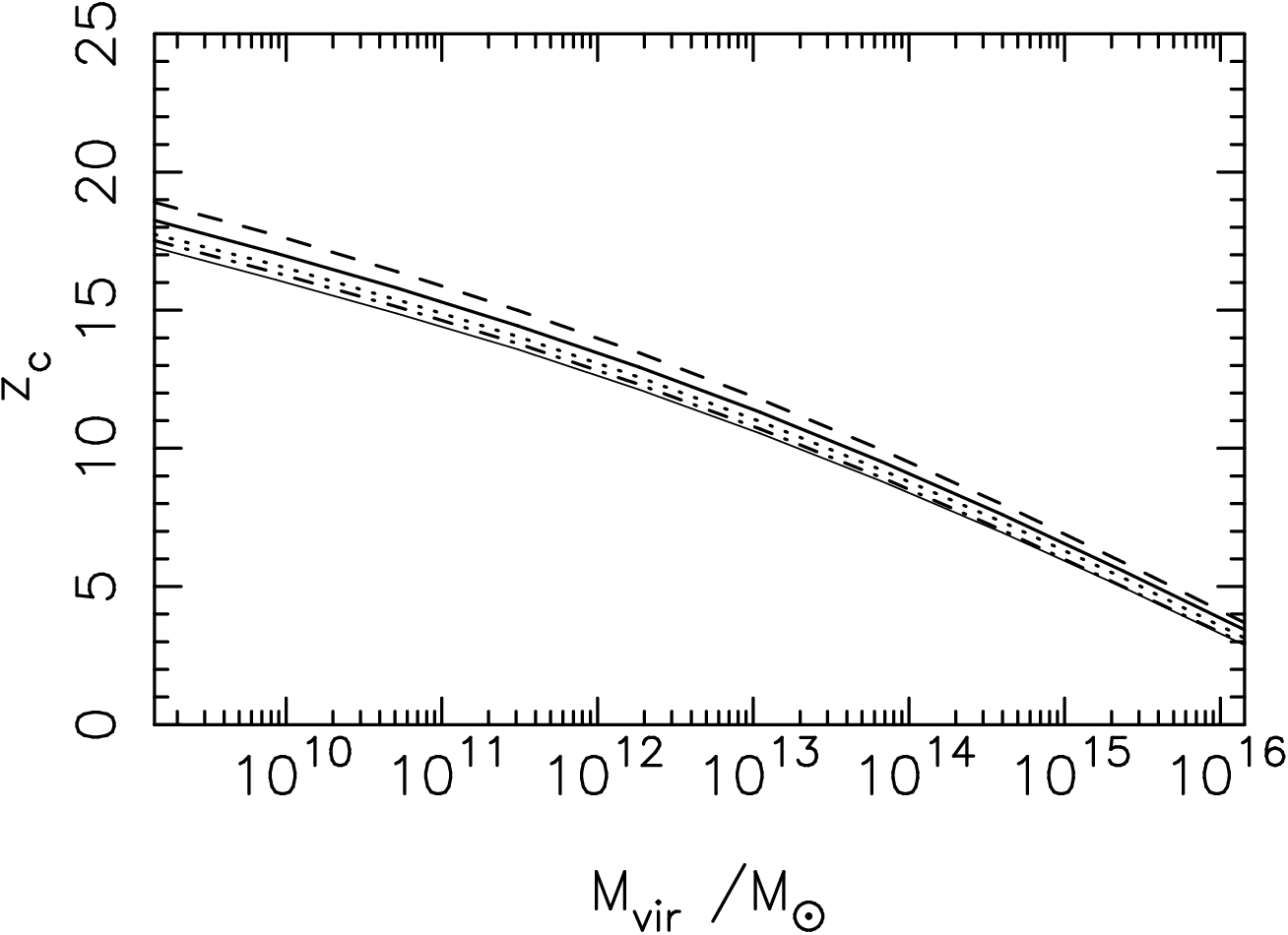}}}
\resizebox{67mm}{!}{
\rotatebox{-90}{
\includegraphics{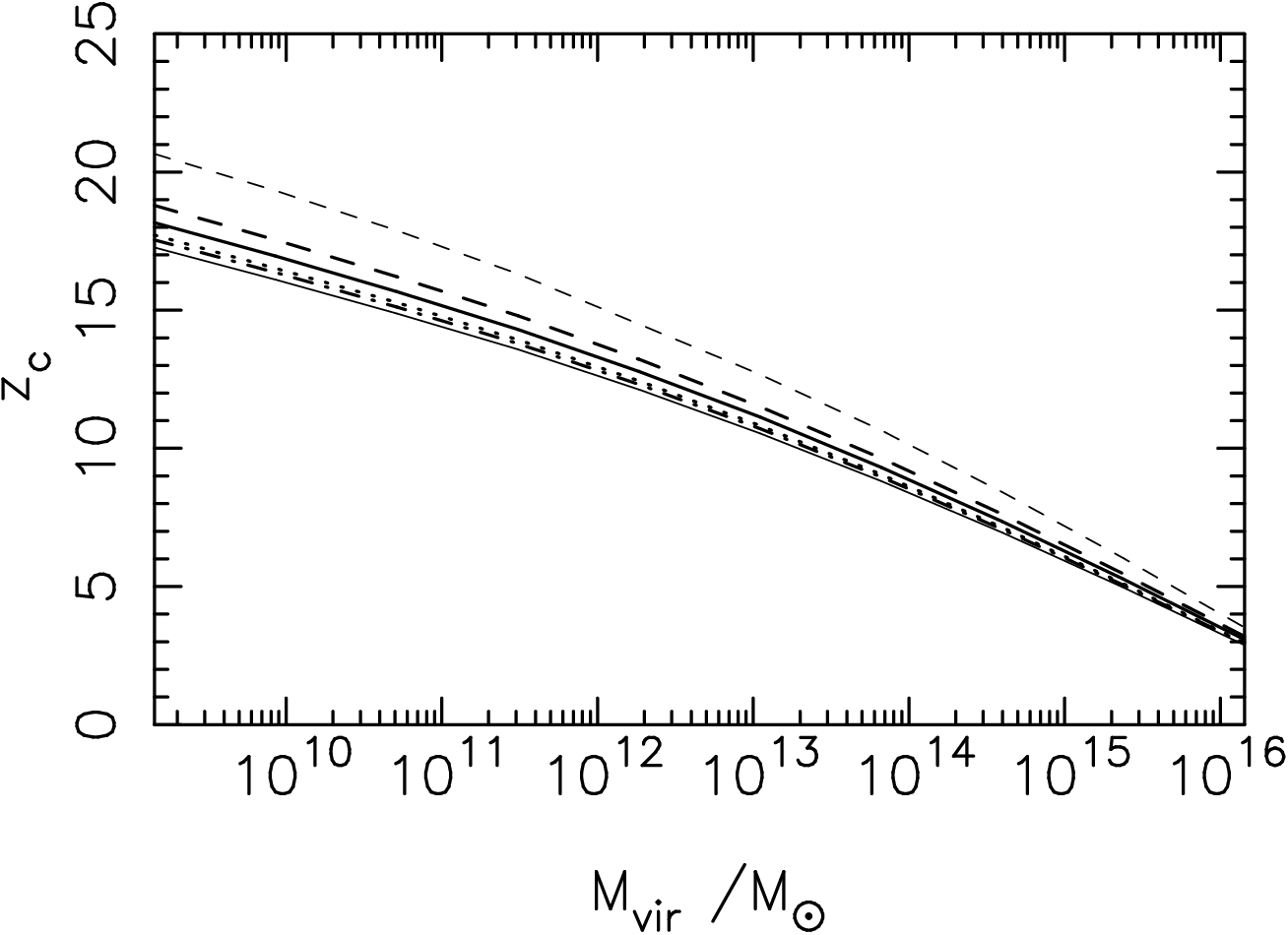}}}
\end{center}
\caption{
{\em left} four inhomogeneous dark energy models (thick solid line -
\cite{steinhardt}, dashed line - \cite{brax}, dot--dashed line -
\cite{skordis}, dotted line - \cite{copel} plus a
cosmological constant model (thin solid line); 
{\em right} six homogeneous dark energy models (thick solid line -
\cite{steinhardt}, thick dashed line - \cite{brax}, dot--dashed line
- \cite{skordis}, dotted line - \cite{copel}, thin dashed line
- constant equation of state $w=-0.6$,  thin solid line - cosmological
constant);  
{\em upper panel} dark matter halo concentrations; 
{\em upper middle panel} non--linear overdensity at virialisation;
{\em lower middle panel} non--linear overdensity at virialisation normalised at $z_{o}$; 
{\em lower panel} characteristic formation redshift as defined by \cite{eke}.}
\label{conc}
\end{minipage}
\end{figure*}
Numerical simulations repeatedly show that the later a halo forms the
less concentrated it is. This appears as a reflection of the smaller
cosmic density at later cosmic epochs. In hierarchical structure
formation models less massive halos form earlier than more massive
ones, hence the former are more concentrated than the
latter.

We use the prescription of \cite{eke} to calculate dark matter halo
concentrations. This prescription consists of a simple analytic
algorithm that was found to reproduce well the mass and redshift
dependence of concentrations obtained from high--resolution N--body
simulations performed in various cosmological models, in particular in
a flat model with a cosmological constant. 

One should, however, point out that the ENS model is by no means the only 
viable prescription. In fact, it is just one of a number of theoretical models for how 
the concentration mass relation arises. Examples of other models are the original NFW model,  the Bullock et al. model  \cite{bullock}, or the recent 
extension of this by Maccio et al. \cite{maccio1}.  Some of these studies produced results in agreement with ENS 
(e.g. \cite{neto}), others find a mass and redshift dependence 
that contradicts the ENS model (e.g. \cite{mac3}, \cite{gao}). At present the matter has not been settled, and it is not 
possible to identify one model that works better than all of the 
others. In this work we will restric ourselves to the ENS model. 
A Further investigation using and comparing other prescriptions will 
appear in a near future work \cite{future}.

The ENS algorithm relates halo
properties with the physics of halo formation as follows:   
\begin{equation}
c_{\rm vir}^{3}=\frac{\Delta_{\rm vir}(z_{c})}{\Delta_{\rm vir}(z_{o})}
\frac{\Omega_{\rm m}(z_{o})}{\Omega_{\rm m}(z_{c})}\left(\frac{1+z_{c}}{1+z_{o}}\right)^{3}  
\label{ccub}
\end{equation}
which results from setting the core density to the spherical top--hat
density at the characteristic formation epoch. $z_{o}$ is the redshift at which the
halo is identified which, is assumed to be the redshift at which the halo as a whole 
collapses and virialises.
The characteristic formation epoch (also designated rapid--collapse
epoch), $z_{c}$, is associated with the collapse time of the halo
subunits. This occurs at the time when the rapid mass accretion rate
drops below a fixed value. Thereafter, the halo evolves and its virial
radius and halo mass grow through minor mergers and diffuse mass
accretion, while  the characteristic length $r_{s}$ remains
essentially equal \cite{kuhlen,wechsler}. This process continues
until the halo as a whole finally collapses and virialises. The
rapid--collapse epoch of a halo of mass $M$ depends on the linear
growth factor, $D(z)$, the amplitude and shape of the matter power
spectrum, and on a single free parameter $C_{\sigma}$ 
whose value can be found which yields concentration values that match
the results from N--body simulations. More explicitly, \cite{eke} define the collapse 
redshift as:
\begin{equation}
D(z_{c})\sigma_{\rm eff}(M_{s})=\frac{1}{C_{\sigma}}
\end{equation}
where $D(z)$ is the linear growth factor and $\sigma_{\rm eff}$ is an
effective amplitude of the power spectrum that \cite{eke} utilise to
modulate $\sigma(M)$ in order to model WDM models. This effective
amplitude is computed at $M_{s}$, the mass enclosed within the radius
at which the NFW circular velocity reaches its maximum,
$r_{\rm max}=2.17r_{s}$. 

\cite{eke} found that
$C_{\sigma}=28$ gives good agreement with the N--body simulations effectuated in  a flat 
cosmological constant model.   
\cite{dolag} have recently performed high--resolution numerical
simulations in dark energy models with constant and time--varying
equation of state (including Ratra--Peebles and SUGRA models) and have
shown that the mass dependence and redshift evolution of
concentrations is in conformity with the \cite{eke} prescription with
$C_{\sigma}=28$. Moreover, these authors have demonstrated for the first
time with numerical simulations that concentrations are indeed larger in dark
energy models as a consequence of halos forming earlier in these
models when compared to the cosmological constant model. This confirms
the predictions of \cite{bart02} and \cite{lopes} which were
based on the analytic algorithm of \cite{eke}

In this work we assume that the \cite{eke} prescription remains valid
for models with inhomogeneous dark energy, once all the cosmological
functions inherent to the algorithm are specified accordingly. 
This may in fact be a strong assumption. In particular, the value assumed for the fitting 
parameter $C_{\sigma}=28$ might not be the correct one. However, this is the
only reference value one has at present. Up until now, there are no
numerical simulations which 
investigate the effects of dark energy on the internal dynamics of the halo
formation. Dark energy has only been incorporated in the background
evolution. Both its pressure and energy density have not even been
considered to contribute to the 
local gravitational potential inside overdensities. 
Although in a very simplified way, the first studies which took into consideration such contributions were the 
N--body simulations performed by \cite{maccio} in the case of coupled quintessence.

For the shape of the linear matter power spectrum we adopt the fitting
formula of \cite{ma} for QCDM models and the \cite{bardeen} fitting
formula for the $\Lambda$CDM model using the modification of \cite{sug} 
in order to account for the baryons. We fix the primordial power 
spectrum index to $n=1$ and the baryon density to $\Omega_{b} h^{2}=0.02$.  
The normalisation of the power spectrum is set by $\sigma_8$, the
$\it{rms}$ linear density fluctuation in spheres of
$8$\,h$^{-1}$\,Mpc, which we choose to be $\sigma_8=0.9$. 

\subsubsection{Inhomogeneous versus homogeneous models} 

Figure \ref{conc} shows halo concentrations for all the models studied
in this paper (on the left - inhomogeneous dark energy models plus a
cosmological constant model for comparison, on the right - homogeneous
dark energy models, including also the cosmological constant model for
comparison). Also shown are the rapid--collapse redshifts, $z_c$ (as defined by \cite{eke}) 
and the non--linear overdensity at virialisation, $\Delta_{\rm vir}$, for the
respective models. 

The upper panel indicates that the \cite{eke} prescription predicts, in general, larger halo concentrations in 
inhomogeneous dark energy models than in homogeneous ones. 
This is however model dependent.
For instance, the \cite{brax} inhomogeneous model is the
model with the highest concentrations which are nearly a factor of two
higher than the cosmological constant model or its homogeneous counterpart. 
In contrast, the inhomogeneous \cite{skordis} model is almost indistinguishable from the cosmological constant model and its  homogeneous counterpart. The inhomogeneous \cite{copel} model
and \cite{steinhardt} model lie in
between the cosmological constant model and the \cite{brax}
inhomogeneous model. The former reveal concentrations higher than their homogeneous counterparts. Furthermore, all homogeneous models present concentrations of the same order of the cosmological constant model.

This is interpreted as a combined effect
arising from the ratios that enter as factors in formula (\ref{ccub}). The matter density ratio is essentially equal in all the models. Although there is some difference
on the rapid--collapse redshifts (lower panel), the overall change in concentrations for inhomogeneous models is primarily
owing to the contrast in non--linear overdensity at virialisation (middle panels). The latter depends on the clustering properties of the quintessence field. While at high virialisation redshifts, all models predict 
$\Delta_{\rm vir} \approx 178$, significant deviations may occur at low virialisation redshifts due to the fact that dark energy starts to dominate in the background universe.
Indeed, for inhomogeneous dark energy ($\Gamma=0$), $\Delta_{\rm vir}$ can differ by a factor of four or more at low virialisation redshifts. This is nevertheless model dependent. In the \cite{skordis} model, the field behaves like a cosmological constant in the background (see Fig \ref{wz}). Hence, one would expect small differences between that and the $\Lambda$CDM model. This is 
indeed the case, because $\dot \phi_{\rm c} \approx 0$ which, in turn, implies $\Gamma \approx 0$, as it can be seen from equation (\ref{gamma}). Thus, in this model, the fluctuations in the quintessence field remain small and the field is almost homogeneous. In fact, the non--linear overdensity for the Albrecht
\& Skordis model oscillates around the cosmological constant model which
explains the virtually coinciding concentrations in these two models. One should
point out that what really characterises the differing ratios
$\Delta_{\rm vir}/\Delta_{\rm vir}(z_{o})$, and consequent varying concentrations in the various models, is the value
of the non--linear overdensity at $z_{o}$ (here, $z_{o}=0.05$) since at
high redshifts, where the rapid--collapse of the halo happens, its value is
practically equal in all models \footnote{it approaches the  Einstein--de--Sitter model $\Delta_{\rm vir}\approx 178$.}. 

In effect, one can see that
$\Delta_{\rm vir}$ at low redshifts can differ by a factor of two between
the cosmological constant model and an inhomogeneous model, but
differs only about ten per cent between the cosmological constant model
and an homogeneous dark energy model. As a result, homogeneous dark
energy models exhibit less noticeable differences in
concentrations (unless one considers high values of $w$) and these are mostly inherent to the differing rapid--collapse redshifts which are highest for the model with constant equation of state $w=-0.6$. Indeed, for homogeneous dark energy models concentrations are larger as $w$ increases owing to structures forming earlier. This is in agreement with the previous findings from high--resolution numerical simulations \cite{klypin,dolag,kuhlen} which confirm that halos keep a memory in their central regions of the mean density of the universe at the characteristic formation epoch.  
Why do all the models, both homogeneous and inhomogeneous models, exhibit  analogous values of $z_c$? It is because $z_c$ is calculated from linear theory, depending solely on the matter power spectrum and on the linear growth factor. Hence, this depends on the background cosmology only and not on the non--linear physics dynamics of the halo that distinctively categorises the inhomogeneous dark energy models and from which one can infer the linear overdensity at collapse. The omission of the latter in the \cite{eke} prescription may be invalid, and thus needs to be tested.

To summarise, the trends presented in Fig. \ref{conc} for halo concentrations in the two classes of models studied appear to be controlled both by differing formation histories as well as by varying virialisation overdensities. While for inhomogeneous models, the virialisation overdensities constitute the dominant factor, for homogeneous models it is the differing formation histories that account for the changes in concentrations. 
 
\section{Observational Implications}
\begin{figure*}
\begin{minipage}{168mm}
\resizebox{80mm}{!}{
\rotatebox{-90}{
\includegraphics{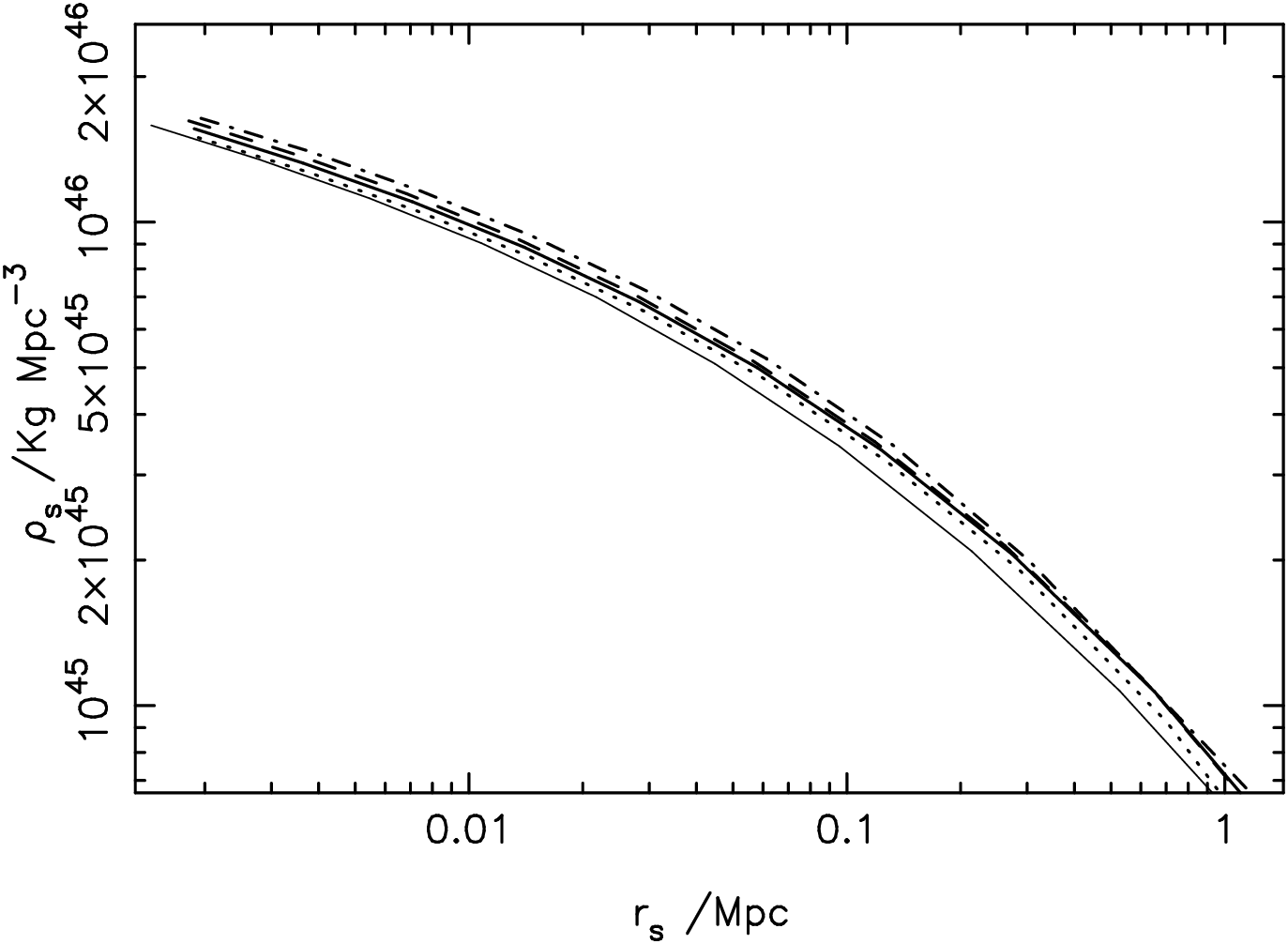}}}
\resizebox{80mm}{!}{
\rotatebox{-90}{
\includegraphics{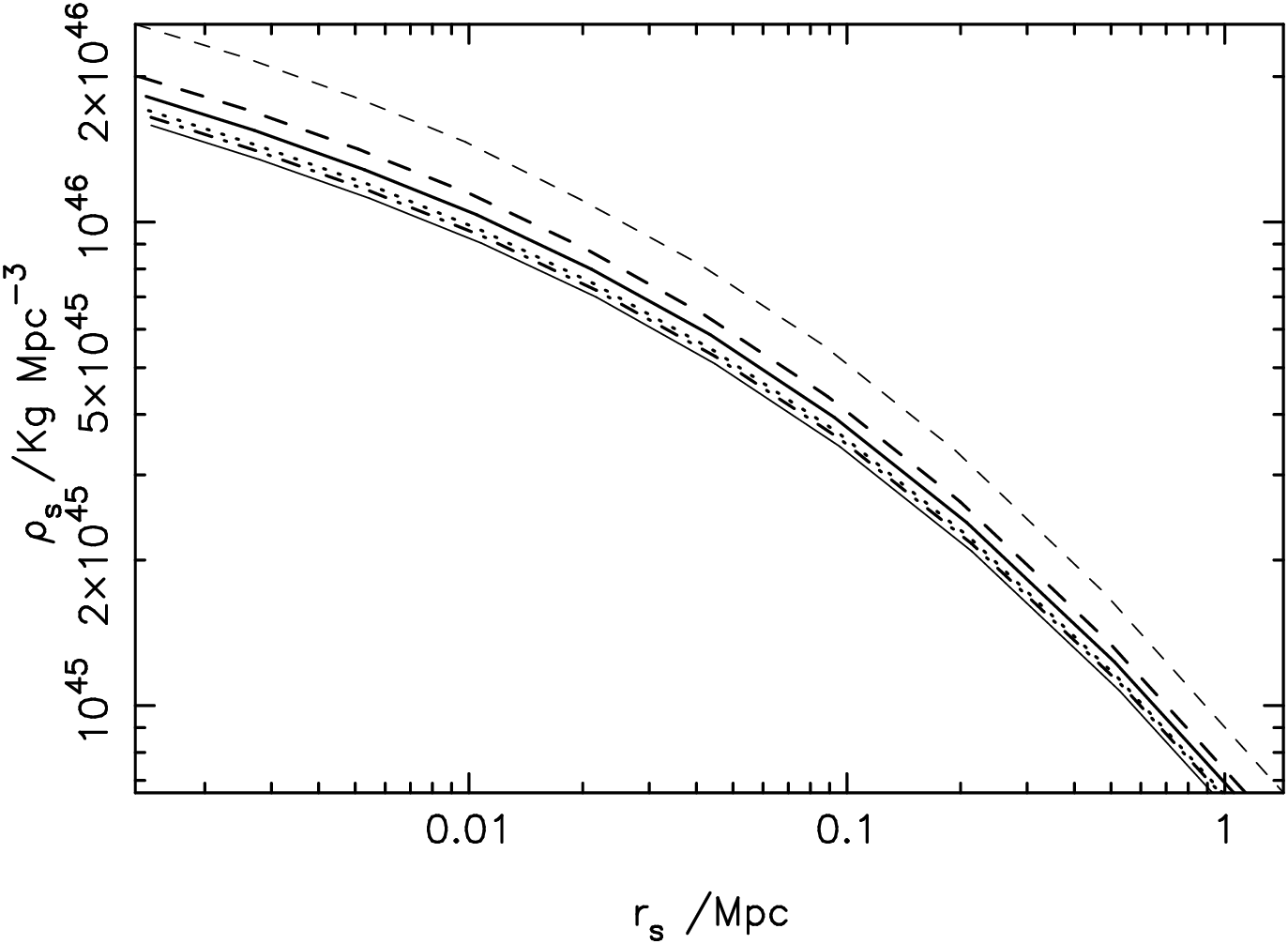}}}
\caption{{\em left} $\rho_{s}$ vs. $r_{s}$ relation in four
  inhomogeneous dark energy models (thick solid line - \cite{steinhardt}, dashed 
line - \cite{brax}, dot--dashed line - \cite{skordis}, dotted line - \cite{copel} 
plus a cosmological constant model (thin solid line); {\em right} the same 
relation in six homogeneous dark energy models (thick solid line - 
\cite{steinhardt}, thick dashed line - \cite{brax}, dot--dashed line - 
\cite{skordis}, dotted line - \cite{copel}, thin dashed line - constant equation 
of state $w=-0.6$,  thin solid line - cosmological constant).} 
\label{xray}
\end{minipage}
\end{figure*}
\begin{figure*}
\begin{minipage}{168mm}
\resizebox{80mm}{!}{
\rotatebox{-90}{
\includegraphics{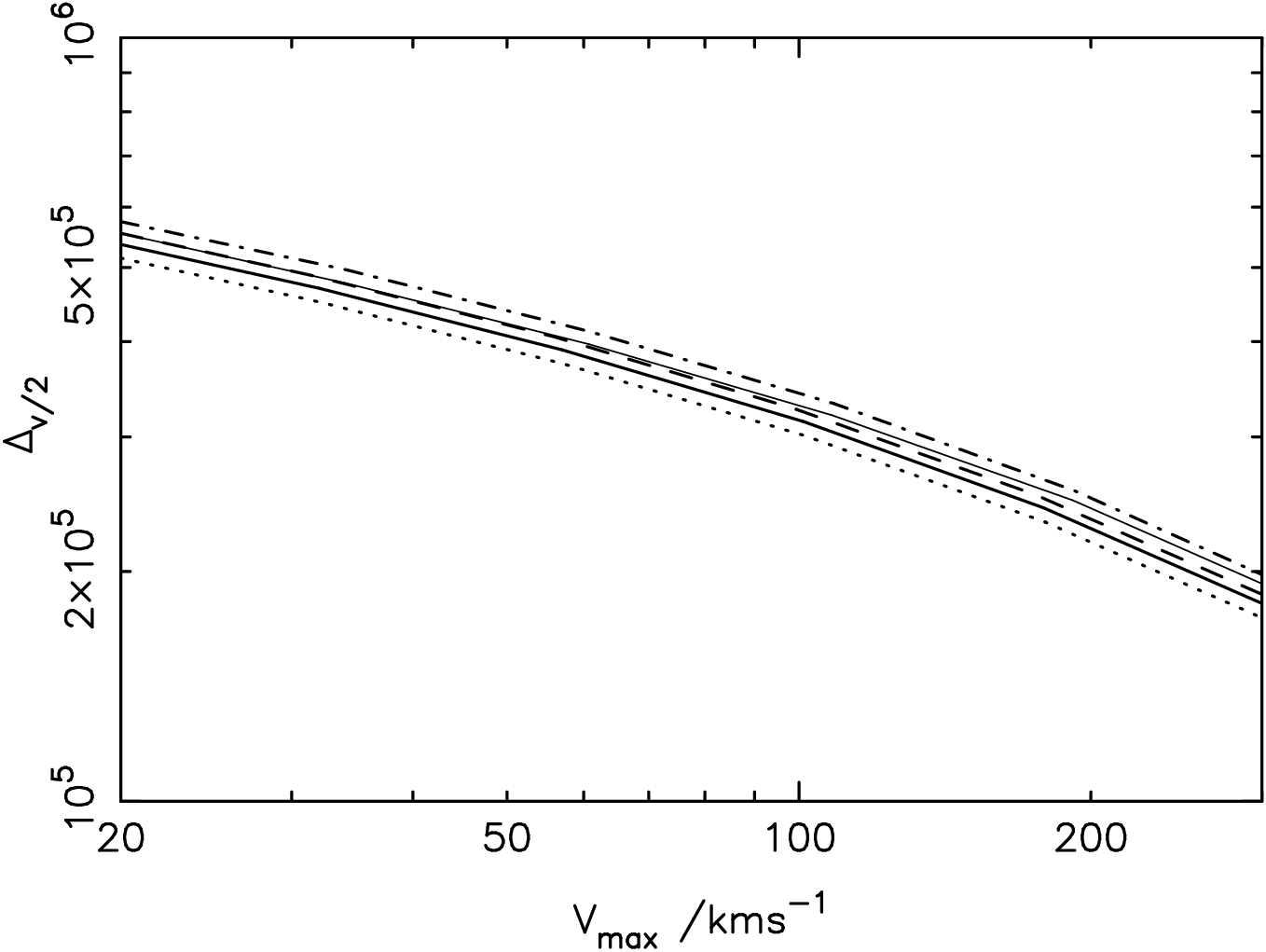}}}
\resizebox{80mm}{!}{
\rotatebox{-90}{
\includegraphics{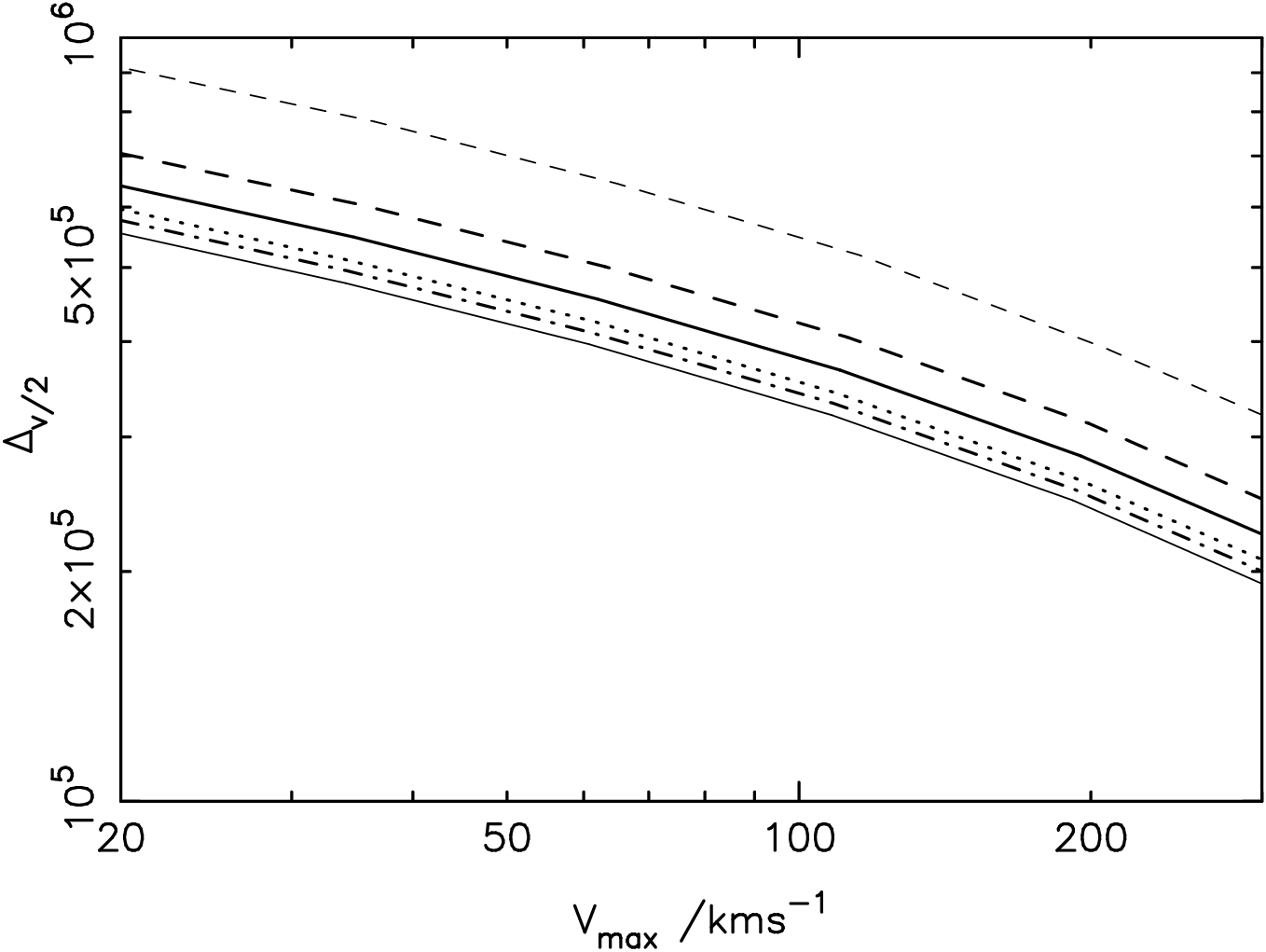}}}
\caption{{\em left} $\Delta_{v/2}$ vs. $V_{\rm max}$ in four 
inhomogeneous dark energy models (thick solid line - \cite{steinhardt}, 
dashed line - \cite{brax}, dot--dashed line - \cite{skordis}, dotted line - 
\cite{copel} plus a cosmological constant model (thin solid line); {\em right} 
the same relation in six homogeneous dark energy models (thick solid line - 
\cite{steinhardt}, thick dashed line - \cite{brax}, dot--dashed line - 
\cite{skordis}, dotted line - \cite{copel}, thin dashed line - constant 
equation of state $w=-0.6$,  thin solid line - cosmological constant).} 
\label{deltav2vsvmax}
\end{minipage}
\end{figure*}
The effect of dark energy on halo properties is expected to be imprinted on 
observable quantities. Unfortunately, the predicted $c_{\rm vir}-M_{\rm vir}$ relation 
is not itself directly observable. 

Note that \cite{cheng} have determined the $c_{\rm vir}-M_{\rm vir}$ relation using the
dynamical properties of dark halos derived from the distribution of
galaxies in massive systems (groups or clusters) and the rotation
curves in less massive systems (dwarf, low surface brightness and
spiral galaxies) and have found that the observational data are
grossly consistent with the theoretical predictions of a flat
cosmological constant model on a mass range $10^{10}-10^{15}
M_{\odot}$. However, measurements of halo concentrations through the 
aforementioned observational method, as well as via weak and/or strong lensing 
cluster measurements, cannot be used to discriminate cosmological models since 
observational estimations of their values rely on $\Delta_{\rm vir}$ and $\rho_{\rm crit}$ 
which are cosmology--dependent but unobservable. 

In the following, we examine two possible methods for comparing observations 
with theoretical predictions. The first method works on galaxy cluster scales 
and consists of fitting the observed X--ray cluster gas density distributions to 
those predicted for an NFW profile. \cite{makino} have
analytically shown that the density distribution of an isothermal gas
cloud with temperature $T_{X}$, in hydrostatic equilibrium, within an
NFW dark matter halo is: 
\begin{equation}
\rho_{gas}(r) = \rho_{gas}(0) e^{-\alpha}(1+r/r_{s})^{\alpha/(r/r_{s})}
\end{equation}
where $\alpha = 4\pi G\mu m_{p} \rho_{s} r_{s}^{2}/kT_{X}$ and $\mu$
and $m_{p}$ designate the mean molecular weight and the proton mass,
respectively.   
The best--fit parameters are thus $\alpha$ and $r_{s}$, or,
alternatively, if one utilises the X--ray temperature, $\rho_{s}$ and $r_{s}$. 
\cite{wu} have applied this technique to an ensemble of 63 X--ray luminous clusters 
and have determined (assuming a flat cosmological constant model) a best--fit 
$c_{\rm vir}-M_{\rm vir}$ relation. Here, we look directly at the quantities $\rho_{s}$ and 
$r_{s}$ which are not dependent on cosmology--dependent but unobservable quantities 
such as $R_{\rm vir}$, although comparison with observations rests on assumptions 
such as the hydrostatic equilibrium and an isothermal temperature profile 
of the intracluster gas.

The second method works on galaxy scales and involves the observational measurement
 of the so--called central density parameter proposed by \cite{alam}. This parameter 
is defined as the mean dark matter overdensity within the radius, $r_{V/2}$, where 
the galaxy rotation curve is one half of its maximum, $V_{\rm max}$:
\begin{equation}
\Delta_{V/2}=\frac{1}{2}\left(\frac{V_{\rm max}}{H_{0} r_{V/2}}\right)^{2}.
\label{deltav2}
\end{equation}
The link between theory and observations can be made through the 
$\Delta_{V/2}-V_{\rm max}$ relation, where $V_{\rm max}$ works as a measure of the 
absolute size of the halo.
As pointed out by \cite{zentner}, $\Delta_{V/2}$ has the advantage of being 
defined without reference to any particular density or velocity profile. 
Theoretically, the maximum velocity of a NFW halo is given by \cite{zentner}:
\begin{equation}
V^{2}_{\rm max} \simeq 0.216 V^{2}_{\rm vir}\frac{c_{\rm vir}}{ln(c_{\rm vir})-c_{\rm vir}/(1+c_{\rm vir})}
\label{vmax}
\end{equation}
where $V_{\rm vir}$ is the virial velocity and $V^{2}_{\rm vir}\equiv GM_{\rm vir}/R_{\rm vir}$. 
In addition, $r_{V/2}\simeq 0.13 r_{s}$ for a NFW profile. Incorporating equation 
\ref{vmax} into equation \ref{deltav2} yields the expected $\Delta_{V/2}$ 
in theory. This can be computed for any cosmological model given $c_{\rm vir}$ 
and $\Delta_{\rm vir}$.
Comparison with observational data relies, however, on the crucial assumption 
that baryons have not substantially modified the dark matter profile. This 
assumption is weakened, but not unambiguously removed, if observations are 
focused on low--surface brightness and dwarf galaxies which are believed to 
be dominated by dark matter. 

We now look at the theoretical predictions from both methods.
Fig. \ref{xray} plots the theoretical relation $\rho_{s}-r_{s}$ expected in the 
various cosmological models for halos at redshift $z_{o}=0.05$. Fig. 
\ref{deltav2vsvmax} presents the theoretical relation $\Delta_{V/2}-V_{\rm max}$ 
expected in the same cosmological models for halos at redshift $z_{o}=0.05$. 
Surprisingly, one observes that, contrary to what one might have expected, 
inhomogeneous models show less pronounced differences in both $\rho_{s}$ and 
$\Delta_{V/2}$ than homogeneous models. Moreover, the inhomogeneous lines, 
in particular, do not reflect the anticipated $c_{\rm vir}$ trends at fixed mass. 
This occurs because $\rho_{s}$ and $\Delta_{V/2}$ depend not only on $c_{\rm vir}$ 
(exactly on $c^{3}_{\rm vir}/(ln(c_{\rm vir})-c_{\rm vir}/(1+c_{\rm vir}))$) but also on 
$\Delta_{\rm vir}(z_{o})$ \footnote{$\rho_{s}$ has an additional dependence 
on $\rho_{\rm crit}$.} which somewhat counteracts with $c_{\rm vir}$. In effect, 
for inhomogeneous models, the non--linear overdensities at virialisation 
decrease as the concentrations increase. The homogeneous models, conversely, 
present differences among themselves that can be about a factor 1.5. In the 
next section we discuss the feasibility of the aforementioned techniques to 
actually discriminate cosmological models.

Finally, the ability of a cluster to lens a background object, a galaxy
or a quasar, depends critically on the halo profile and consequently
on its concentration, therefore the influence of dark energy (and its 
homogeneity nature) on halo concentrations may be tested, albeit indirectly, 
through the observational statistics 
of multiply imaged quasars or giant arcs of galaxies. \cite{lopes} have 
shown that the cross--section for quasar multiple imaging is increased for 
homogeneous dark energy models with $w > -1$, owing to halo concentrations being
larger in those models. The expected increase on halo concentrations
in models where dark energy is inhomogeneous may also produce an effect on the
lensing cross--section. This remains to be investigated.

\section{Discussion and Conclusions}

We have studied halo properties in models with dynamical dark
energy. This work extends upon previous studies in that we investigate
halo properties as a function of a time--varying dark energy equation
of state, covering four classes of potentials, and its homogeneity
nature in the highly non--linear regime.   

The dynamics of the collapse of the halo is regulated by the form of
the dark energy potential, the time evolution of its equation of
state, the initial conditions of the field and its homogeneity
properties in the highly non--linear regime. These have a direct
influence on the turnaround, virialisation and collapse times,
altering in consequence the (non--linear) density contrast and virial
radius \cite{mota,maor}.  

As for the homogeneity nature of dark energy in the highly non--linear
regime, we have examined two extreme scenarios: first, we assumed that
the quintessence field does not exhibit fluctuations on cluster scales
and below - homogeneous fluid; second, we supposed that the field
inside the overdensity collapses along with the dark matter -
inhomogeneous fluid. 

We have computed halo concentrations using the algorithm of \cite{eke}
and have derived physical halo properties expected within this
analytical treatment. 
We find that the \cite{eke} prescription displays, in general (the exception 
being the \cite{skordis} model), higher halo concentrations in inhomogeneous 
dark energy models than in their homogeneous equivalents. The 
\cite{brax} inhomogeneous model is the model with the highest concentrations 
which are nearly a factor of two higher than the cosmological constant model 
or its homogeneous counterpart.
For homogeneous dark energy models concentrations are
larger as $w$ increases owing to structures forming earlier. This is
in agreement with previous findings from high--resolution numerical
simulations which demonstrate that halos keep a memory in their
central regions of the mean density of the universe at their
characteristic formation epoch.

In the two cases analysed (homogeneous and inhomogeneous models), halo
concentrations seem to be controlled by both changes in formation
epochs of the halo cores as well as by differing virialisation
overdensities. While for inhomogeneous models, changes in the
virialisation process constitute the most influential factor, for
homogeneous models it is the differing formation histories that are
most responsible for the changes in concentrations. 

Having determined the theoretical $c_{\rm vir}-M_{\rm vir}$ relation, we
then deduced the corresponding $\rho_{s}-r_{s}$ and
$\Delta_{V/2}-V_{\rm max}$ relations which represent physical measures
that, unlike the $c_{\rm vir}-M_{\rm vir}$ relation, may establish a
more direct link with observations. Here, we note that the
homogeneous models manifest more detectable differences in both
$\rho_{s}$ and $\Delta_{V/2}$ than the inhomogeneous
models. Specially, the inhomogeneous curves do not register the scaling
one might have expected based on the $c_{\rm vir}$ values at fixed
mass. This arises because $\rho_{s}$ and $\Delta_{V/2}$ depend both on
$c_{\rm vir}$ and on $\Delta_{\rm vir}(z_{o})$ which counterbalance each
other. Nonetheless, it is intriguing why inhomogeneous models reveal
visible changes in their collapse dynamics and do not show apparent
differences in their central densities. This could be associated with
the prescription used to calculate the characteristic collapse epoch
which may not be applicable to inhomogeneous models. 
We notice that it is also important to realise that the free
parameter $C_{\sigma}=28$, whose meaning is in fact unknown, may not
be appropriate. Clearly, these assumptions should not be undervalued,
and require a test against N--body simulations.

Studying the impact of dark energy on the density structure of dark
matter halos surely represents an important step on our understanding
of structure formation on those models. Unfortunately, halo properties
do not seem to provide a (at present) robust way of probing dark
energy. The $\Delta_{V/2}-V_{\rm max}$ and $\rho_{s}-r_{s}$ relations are
plagued by observational scatter on the data that may (or may not) be
associated with selection effects (cooling baryons in the former;
hydrostatic equilibrium and isothermal hypothesis in the latter)
and by the intrinsic, theoretical in nature, scatter in halo
concentration values about the mean which is believed to be due to a
spread on the collapse histories of N--body simulated
halos. Additionally, there is the implicit uncertainty on the the
value of $\sigma_8$ as a function of $w$ which we fixed in this
analysis (see discussion in \cite{kuhlen}).  
Nevertheless, the effect of dark energy on the dark matter halo
structure may be exploited indirectly through strong and weak lensing
statistics. Another alternative path may be the time evolution of dark
matter halo abundances \cite{klypin}. 

While it is reassuring that the analytical recipe of \cite{eke} is
successful in reproducing halo concentrations measured in numerical
simulations performed in homogeneous dark energy models, more
explicitly for SUGRA and Ratra--Peebles potentials \cite{dolag}, it
should be recognised that the same recipe has not been tested against
simulations carried out in inhomogeneous models.  
Moreover, one should point out that the methods here used are based on the premise that the halo concentration-mass relation is 
governed by the ENS  model. However,  the ENS model is by no means the only 
viable and popular model. It is just one of a number of theoretical models for how 
the concentration - mass relation arises. See \cite{nfw,bullock,maccio1} for other examples of other possibilities.
Furthermore, since the Dolag et al. paper \cite{dolag} (the most recent 
paper advocating the ENS prescription) there have 
been a number of developments in studying the halo concentration - mass - redshift 
relation. Some of these studies produced results in agreement with ENS 
(e.g. Neto et al. \cite{neto}), others find a mass and redshift dependence 
that contradicts the ENS model (e.g. Maccio et al. \cite{mac3} and Gao et 
al. \cite{gao}). At present the matter has not been settled, and it is not 
possible to identify one model that works better than all of the 
others. A Further investigation using and comparing other prescriptions will appear in a near future work \cite{future}.

Ideally, this work will motivate further studies using N--body
simulations which, ultimately, will draw a firm conclusion on the
interconnection between the dark matter halo structure and the
underlying cosmology.

\section*{Acknowledgments} 

We would like to thank C. van de Bruck, A. Lopes, L. Miller and J. Taylor for useful discussions and the referee for useful comments. 
DFM acknowledges support from the A. Humboldt Foundation and the CNRS Visitor Research Fellowship at the University of Montpellier. 

\section*{References} 

\end{document}